\documentclass[twocolumn]{aastex631}
\shorttitle{Search for galaxy candidates at z$\sim$9-15 in GLASS-JWST--ERS}
\shortauthors{Marco Castellano and the GLASS-JWST team}

\usepackage{hyperref}  
\usepackage{amsmath} 
\usepackage{rotating,tabularx}
\hypersetup{colorlinks=true,linkcolor=[rgb]{1.,0.2,0.2},citecolor=[rgb]{0.1,0.4,1.},filecolor=[rgb]{0.7,0.2,0.2},urlcolor=[rgb]{0.7,0.2,0.2}}

\usepackage{color}
\usepackage{makecell}
\usepackage{graphicx}
\usepackage{multirow}
\usepackage{txfonts}
\definecolor{blue}{rgb}{0., 0., 1}

\begin{document}

\title{Early results from GLASS-JWST. III: Galaxy candidates  at z$\sim$9-15\footnote{Based on observations collected with JWST under the ERS programme ID 1324 (PI T. Treu)}}

\correspondingauthor{Marco Castellano}
\email{marco.castellano@inaf.it}

\author[0000-0001-9875-8263]{Marco Castellano}
\affiliation{INAF - Osservatorio Astronomico di Roma, via di Frascati 33, 00078 Monte Porzio Catone, Italy}

\author[0000-0003-3820-2823]{Adriano Fontana}
\affiliation{INAF - Osservatorio Astronomico di Roma, via di Frascati 33, 00078 Monte Porzio Catone, Italy}

\author[0000-0002-8460-0390]{Tommaso Treu}
\affiliation{Department of Physics and Astronomy, University of California, Los Angeles, 430 Portola Plaza, Los Angeles, CA 90095, USA}

\author[0000-0002-9334-8705]{Paola Santini}
\affiliation{INAF - Osservatorio Astronomico di Roma, via di Frascati 33, 00078 Monte Porzio Catone, Italy}

\author[0000-0001-6870-8900]{Emiliano Merlin}
\affiliation{INAF - Osservatorio Astronomico di Roma, via di Frascati 33, 00078 Monte Porzio Catone, Italy}

\author[0000-0003-4570-3159]{Nicha Leethochawalit}
\affiliation{School of Physics, University of Melbourne, Parkville 3010, VIC, Australia}
\affiliation{ARC Centre of Excellence for All Sky Astrophysics in 3 Dimensions (ASTRO 3D), Australia}
\affiliation{National Astronomical Research Institute of Thailand (NARIT), Mae Rim, Chiang Mai, 50180, Thailand}

\author[0000-0001-9391-305X]{Michele Trenti}
\affiliation{School of Physics, University of Melbourne, Parkville 3010, VIC, Australia}
\affiliation{ARC Centre of Excellence for All Sky Astrophysics in 3 Dimensions (ASTRO 3D), Australia}

\author[0000-0002-5057-135X]{Eros~Vanzella}
\affiliation{INAF -- OAS, Osservatorio di Astrofisica e Scienza dello Spazio di Bologna, via Gobetti 93/3, I-40129 Bologna, Italy}

\author[0000-0002-0441-8629]{Uros Mestric}
\affiliation{INAF -- OAS, Osservatorio di Astrofisica e Scienza dello Spazio di Bologna, via Gobetti 93/3, I-40129 Bologna, Italy}

\author{Andrea Bonchi}
\affiliation{Space Science Data Center, Italian Space Agency, via del Politecnico, 00133, Roma, Italy}

\author{Davide Belfiori}
\affiliation{INAF - Osservatorio Astronomico di Roma, via di Frascati 33, 00078 Monte Porzio Catone, Italy}

\author[0000-0001-6342-9662]{Mario Nonino}
\affiliation{INAF - Osservatorio Astronomico di Trieste, Via Tiepolo 11, I-34131 Trieste, Italy}

\author{Diego Paris}
\affiliation{INAF - Osservatorio Astronomico di Roma, via di Frascati 33, 00078 Monte Porzio Catone, Italy}

\author{Gianluca Polenta}
\affiliation{Space Science Data Center, Italian Space Agency, via del Politecnico, 00133, Roma, Italy}

\author[0000-0002-4140-1367]{Guido Roberts-Borsani}
\affiliation{Department of Physics and Astronomy, University of California, Los Angeles, 430 Portola Plaza, Los Angeles, CA 90095, USA}

\author[0000-0003-4109-304X]{Kristan Boyett}
\affiliation{School of Physics, University of Melbourne, Parkville 3010, VIC, Australia}
\affiliation{ARC Centre of Excellence for All Sky Astrophysics in 3 Dimensions (ASTRO 3D), Australia}

\author[0000-0001-5984-0395]{Marusa Bradac}
\affiliation{University of Ljubljana, Department of Mathematics and Physics, Jadranska ulica 19, SI-1000 Ljubljana, Slovenia}
\affiliation{Department of Physics and Astronomy, University of California Davis, 1 Shields Avenue, Davis, CA 95616, USA}

\author[0000-0003-2536-1614]{Antonello Calabr\`o}
\affiliation{INAF - Osservatorio Astronomico di Roma, via di Frascati 33, 00078 Monte Porzio Catone, Italy}

\author[0000-0002-3254-9044]{K. Glazebrook}
\affiliation{Centre for Astrophysics and Supercomputing, Swinburne University of Technology, PO Box 218, Hawthorn, VIC 3122, Australia}

\author[0000-0002-5926-7143]{Claudio Grillo}
\affiliation{Dipartimento di Fisica, Università degli Studi di Milano, via Celoria 16, I-20133 Milano, Italy}
\affiliation{INAF—IASF Milano, via A. Corti 12, I-20133 Milano, Italy}

\author[0000-0002-9572-7813]{Sara Mascia}
\affiliation{INAF - Osservatorio Astronomico di Roma, via di Frascati 33, 00078 Monte Porzio Catone, Italy}

\author[0000-0002-3407-1785]{Charlotte Mason}
\affiliation{Cosmic Dawn Center (DAWN)}
\affiliation{Niels Bohr Institute, University of Copenhagen, Jagtvej 128, 2200 København N, Denmark}

\author[0000-0001-9261-7849]{Amata Mercurio}
\affiliation{INAF -- Osservatorio Astronomico di Capodimonte, Via Moiariello 16, I-80131 Napoli, Italy}

\author[0000-0002-8512-1404]{T. Morishita}
\affiliation{IPAC, California Institute of Technology, MC 314-6, 1200 E. California Boulevard, Pasadena, CA 91125, USA}

\author[0000-0003-2804-0648 ]{Themiya Nanayakkara}
\affiliation{Centre for Astrophysics and Supercomputing, Swinburne University of Technology, PO Box 218, Hawthorn, VIC 3122, Australia}

\author[0000-0001-8940-6768 ]{Laura Pentericci}
\affiliation{INAF - Osservatorio Astronomico di Roma, via di Frascati 33, 00078 Monte Porzio Catone, Italy}

\author[0000-0002-6813-0632]{Piero Rosati}
\affiliation{Dipartimento di Fisica e Scienze della Terra, Università degli Studi di Ferrara, Via Saragat 1, I-44122 Ferrara, Italy}
\affiliation{INAF - OAS, Osservatorio di Astrofisica e Scienza dello Spazio di Bologna, via Gobetti 93/3, I-40129 Bologna, Italy}

\author[0000-0003-0980-1499]{Benedetta Vulcani}
\affiliation{INAF Osservatorio Astronomico di Padova, vicolo dell'Osservatorio 5, 35122 Padova, Italy}

\author[0000-0002-9373-3865]{Xin Wang}
\affil{Infrared Processing and Analysis Center, Caltech, 1200 E. California Blvd., Pasadena, CA 91125, USA}

\author[0000-0002-8434-880X]{L.~Yang}
\affiliation{Kavli Institute for the Physics and Mathematics of the Universe, The University of Tokyo, Kashiwa, Japan 277-8583}


%
%
%





\begin{abstract}
We present the results of a first search for galaxy candidates at z$\sim$9--15 
on deep seven-bands NIRCam imaging acquired as part of the GLASS-JWST Early Release Science Program on a flanking field of the Frontier Fields cluster A2744. Candidates are selected via two different renditions of the Lyman-break technique, isolating objects at z$\sim$9-11, and z$\sim$9-15, respectively, supplemented by photometric redshifts obtained with two independent codes. We find five color-selected candidates at  z$>$9, plus one additional candidate with photometric redshift z$_{phot}\geq$9. 
In particular, we identify two bright candidates at $M_{UV}\simeq -21$ that are unambiguously placed at $z\simeq 10.6$ and $z\simeq 12.2$, respectively. The total number of galaxies discovered at $z>9$ is in line with the predictions of a non-evolving LF. The two bright ones at $z>10$ are unexpected given the survey volume, although cosmic variance and small number statistics limits general conclusions. This first search demonstrates the unique power of JWST to discover galaxies at the high redshift frontier. The candidates are ideal targets for spectroscopic follow-up in Cycle$-2$.
\end{abstract}




\section{Introduction}\label{sec:intro}
Since the discovery of the first object at a cosmological distance \citep[][]{Schmidt1963}, the quest for ``the most distant source ever seen'' has been both a motivation and a powerful mean to expand our understanding of cosmology and of how astronomical objects formed and evolved. Substantial progress in the study of distant, star-forming galaxies has been enabled by the Lyman-break technique yielding to samples of so-called Lyman-break galaxies (LBGs) at progressively higher redshifts \citep[e.g.][]{Giavalisco2002}. 

In recent years, the combination of Hubble Space Telescope (HST) and ground-based data enabled the selection of relatively large samples of LBGs at z$\sim$2--8 \citep[e.g.][]{Reddy2009,McLure2013,Finkelstein2015,Roberts-Borsani2016,Ono2018,Atek2018}. The inferred UV luminosity functions (LF), together with assumptions on the ionizing output of LBGs, enabled the first investigations of the epoch of reionization \citep[EoR, e.g.,][]{Dayal2018}. The data are consistent with a scenario in which the Universe was still largely neutral at z$\sim$8 and was rapidly reionized by $z\sim5.5-6$ \citep[e.g.][]{Mitra2015,Greig2017,Mason2018}, likely by faint star-forming galaxies \citep[e.g.][]{Castellano2016c,Yue2018,Finkelstein2019,Dayal2020}. 

Very few constraints are available at higher redshifts. Small samples of increasingly bright objects have been selected up to z$\sim$9--11 \citep[][]{Zheng2012,Oesch2013,Coe2013,McLeod2016,Hashimoto2018,Ishigaki2018,Morishita2018,Rojas-Ruiz2020,Bouwens2021,Roberts-Borsani2022,Bagley2022}. The available estimates of the Schechter UV LF show significant discrepancies, with some results in agreement with a smooth evolution from lower redshifts, driven either by a decrease of $\phi^*$ or by a dimming of $M^*$ \citep[e.g.][]{McLeod2016}, other analysis suggesting either a dearth \citep[][]{Oesch2018} or an excess \citep[][]{Bagley2022} of bright galaxies at z$\sim$10. As a result,  the cosmic star-formation rate density (SFRD) is poorly constrained, and it is not clear whether the ratio between star-formation rate and dark matter accretion rate remains approximately constant at z$\sim$9-11 as found at z$\lesssim$7 \citep[][]{Harikane2022}.

The discovery of the population of galaxies that ended the so-called ``dark ages'' at z$\sim$12 and beyond has been so far prevented by the limited IR spectral coverage of HST, and by the insufficient sensitivity and resolution of Spitzer and ground-based telescopes. In fact, at such high-redshifts only tentative candidates have been found \citep{Harikane2022}. The highest spectroscopically confirmed redshift is that of LBG GN-z11 at $z=10.957$ \citep[][]{Oesch2016,Jiang2021}. Even with their limitations, current observations have challenged our understanding of galaxy build-up in the early universe. The number densities of bright $z>10$ candidates and GN-z11 lie above the typical empirical Schechter function \citep[e.g.,][]{Bowler2020} while theoretical models are in tension with the abundance of bright galaxies \citep[][]{Finkelstein2022}. Thus, these remarkably bright galaxies may mark a significant transition in the mode of galaxy formation in the early universe: likely significantly less dust-obscured and with a higher star formation efficiency enabling more stars to form in high mass halos, and/or the presence of young quasars. Understanding how these galaxies formed requires building a census of the galaxy population at $z>10$ with deeper photometric data, especially at longer wavelengths than those covered by HST.

In this paper we present the search for LBG candidates at z$\sim$9--15 in the first NIRCam imaging data from the GLASS-JWST Early Release Science Program \citep[JWST-ERS-1324,][T22 hereafter]{TreuGlass22}. GLASS-JWST targets the HFF galaxy cluster A2744 \citep[][]{Lotz2014,Merlin2016b,Castellano2016b} with NIRISS and NIRSpec spectroscopy, while obtaining coordinated parallel observations with NIRCam on two flanking fields. The GLASS NIRCam parallels are the deepest images collected by the 13 ERS programs. Details on the design and strategy of the GLASS-JWST survey can be found in T22. 

Throughout the paper we adopt AB magnitudes \citep{Oke1983}, and a flat $\Lambda$-CDM concordance model ($H_0$ = 70.0 km s$^{-1}$ Mpc$^{-1}$, $\Omega_M=0.30$).

\section{GLASS-JWST NIRCam imaging} \label{sec:data}

The GLASS-JWST NIRCam observations discussed in this paper were taken in parallel to NIRISS on June 28-29 2022, They are centered at RA$=3.5017025$ deg and Dec$=-30.3375436$ deg, and consist of images in seven bands: F090W (total exposure time: 11520 seconds), F115W (11520 s.), F150W (6120 s.), F200W (5400 s.), F277W (5400 s.), F356W (6120 s.), and F444W (23400 s.). The image reduction and calibration, and the methods used to detect sources and measure multi-band photometry are described in a companion paper \citep[paper II in this series,][M22 hereafter]{Merlin2022}. The NIRISS observations are described in paper I \citep[][]{Roberts-Borsani2022b}.

For convenience of the reader, we briefly summarise below the information relevant for the present paper. Data reduction and flux calibration  were obtained using the official JWST pipeline\footnote{\url{https://jwst-docs.stsci.edu/jwst-science-calibration-pipeline-overview}} and exploiting the calibration files made available by STScI on July 29 2022.
A customized step has been introduced to improve masking of image defects ("snowballs" and "wisps") and to improve the subtraction of 1/f noise and background. Images have been aligned to a common reference grid (defined on Gaia DR3 calibrated ground-based images) using a custom pipeline already adopted in similar projects \citep[]{Fontana2014}. 

We warn the reader that flat--fielding, flux calibration and other steps of the data processing are inevitably based on preliminary calibrations. Similarly, the cleaning of defects, especially in the bluest bands that are used here to identify the Lyman break, will also be improved as the knowledge of the telescope and instrument improves.

Objects were detected using a customized version of \textsc{SExtractor} \citep[][v2.8.6]{Bertin1996,Guo2013} on the F444W coadded image. Total F444W fluxes in elliptical apertures as defined by \citet[][]{Kron1980} were measured with \textsc{a-phot} \citep[][]{Merlin2019}. Fluxes in the other bands were measured with \textsc{a-phot} in several apertures at the positions of the detected sources using images PSF-matched to the F444W one. For the final search of high-z candidates we have used both an aperture of 2FWHM diameter (=$0\farcs28$) as well as a smaller, $0\farcs2$ one. The latter has been adopted to optimize the search of potentially blended sources, which may be useful in our crowded field. 

The final 5$\sigma$ depths for point sources within a circular aperture of diameter of $0\farcs2$ are 28.78, 29.03, 28.84, 28.89, 29.26, 29.33, 29.71 in the F090W-F444W bands, respectively (M22). 

We estimate a total area of $\sim$7 sq. arcmin available for the selection of high-redshift candidates by considering the regions observed in all bands, and excluding the pixels occupied by foreground sources. Modest lensing magnification is expected to be present in the NIRCam parallel fields. In this initial set of papers we neglect the effect. The issue will be revisited after the completion of the campaign, which includes a second NIRCam field, in parallel to NIRSpec observations (T22).

\section{Selection of Lyman-break galaxies at z$>$9 with NIRCam}\label{Sect:colorsel}
\subsection{Color-color selection}

We defined color selection criteria for galaxies at z$\sim$9-15 using a  catalogue of objects at z$=$0-15 that mimicks our observations. The catalogue comprises objects at 0$<$z$\leq$4.5 generated with the EGG code \citep{Schreiber2017EGG}, that is based on empirical relations to  reproduce observed number counts and color distributions of galaxies at low and intermediate redshifts. Galaxies at higher redshifts have been generated according to observed UV LFs over z$\sim$5-10 \citep{McLeod2015,Oesch2018,Bouwens2021} and randomly associating to each object a template from a library based on \citet[][BC03 hereafter]{Bruzual2003} models with metallicity=0.02,0.2 Z$_{\odot}$, 0$<$E(B-V)$<$0.2, constant SFH, \citet{Salpeter1955} IMF and \citet{Calzetti2000} extinction law to predict photometry in 7 bands. 

In order to provide sufficient statistics to design appropriate selection criteria, we simulated the catalog over an area of $\sim$0.12 sq. deg., which is $\sim$60 times larger than the NIRCam parallel field, and we assumed no evolution of the UV LF beyond $z=10$. We also artificially boosted the number counts at z$>9.5$ by a factor of 20. The over-representation of high-z sources in the mock catalogs is taken into account when considering purity and completeness. Late type dwarf stars are another potential source of contamination for the color selection of high-redshift galaxies. We assessed their impact using synthetic JWST photometry for the models by \citet[][]{Marley2021} which include brown dwarfs and self-luminous extrasolar planets with $200 \leq T_{eff} \leq 2400$ and metallicity [M/H] from 0.5 to + 0.5. The brown dwarf models were normalized at 26.0$\leq$F444W$\leq$28.0 in 0.5 mag steps.

Finally, the catalogues were perturbed by adding noise in order to reproduce the expected relation and scatter between magnitudes and errors in each band.

\begin{figure}[ht!]
\includegraphics[width=10cm]{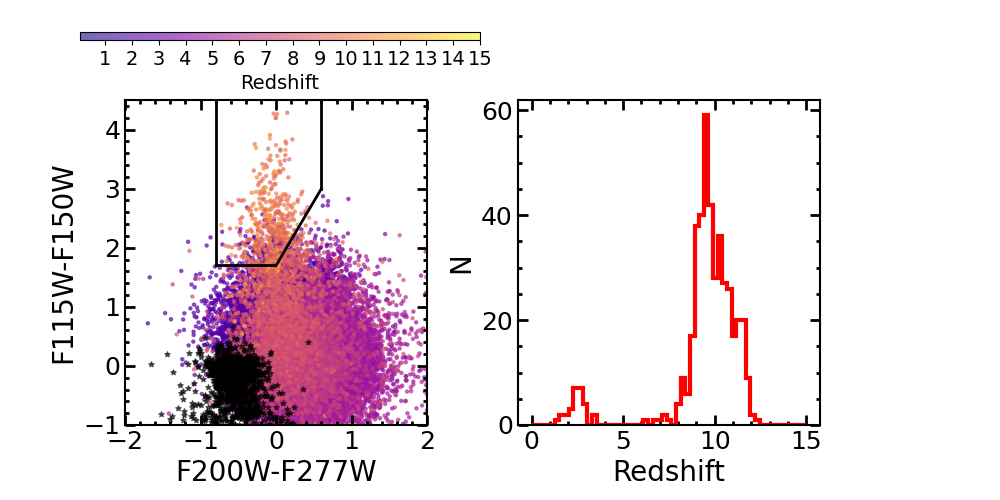}
\includegraphics[width=10cm]{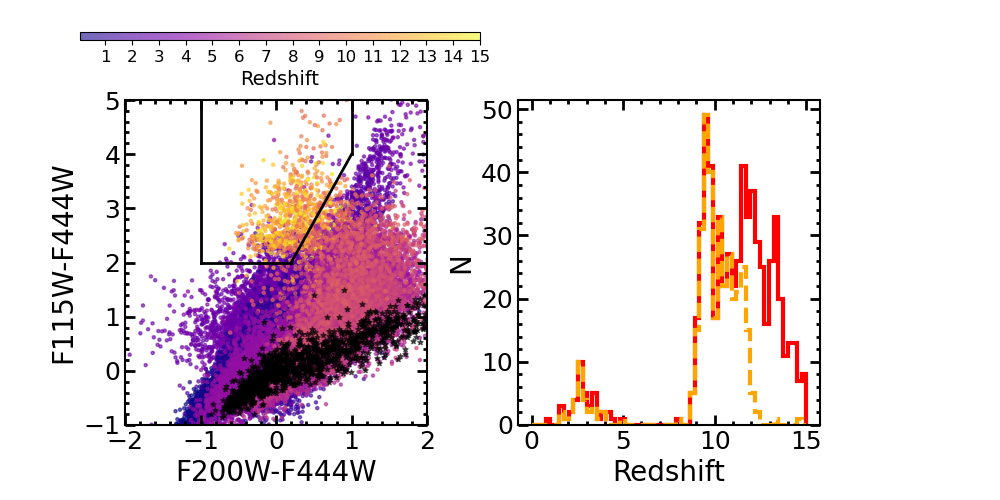}
\caption{Color selection diagrams (left panels), and redshift distribution of selected galaxies (right panels), for LBGs at z$\sim$9-11 (top, Selection \#1) and z$\sim$9-15 (bottom, Selection \#2) exploiting the NIRCam filters of the GLASS-JWST-ERS parallel. The catalog comprises objects at $z=0-15$ over an area of 0.12 sq. deg. Black stars show the position of brown dwarf models from \citet[][]{Marley2021} normalized at 26.0$\leq$F444W$\leq$28.0 that meet our detection criteria for high-redshift galaxies. All fluxes have been perturbed with realistic noise properties.\label{figcolsel}}
\end{figure}

We find that the following color-color selection window, which is a modified version of the selection proposed by \citet[][]{Hainline2020}, identifies objects at z$\sim$9-11.5, as shown also in the top panels of Fig.~\ref{figcolsel}:

\begin{equation}\label{col1}
\begin{aligned}
    &(F115W-F150W)>1.7\\
    &(F115W-F150W)>2.17 \times (F200W-F277W)+1.7\\
    &-0.8<(F200W-F270W)<0.6.
\end{aligned}
\end{equation}

In addition, we require a signal-to-noise ratio SNR$>$3 in all bands redward of the Lyman break (F150W, F200W, F277W, F356W), and SNR$<$1.5 in the F090W band. 

Objects in the wider range z$\sim$9-15 are well identified by the following selection criteria, which leverages also on the redder bands (e.g. F444W):

\begin{equation}\label{col2}
\begin{aligned}
    &(F115W-F444W)>2.0\\
    &(F115W-F444W)>2.5 \times (F200W-F444W)+1.5\\
    &-1.0<(F200W-F444W)<1.0.
\end{aligned}
\end{equation}

By requiring all the SNR requirements above, plus SNR(F150W)$<$2 and SNR(F115W)$<$2 , the selection in Eq.~\ref{col2} proves to be complementary to the one in Eq.~\ref{col1}, sampling the z$\sim$11-15 redshift range (Fig.~\ref{figcolsel}, bottom panels). We verified that most of the z$\lesssim$11 objects selected in the F115W-F150W vs F200W-F277W diagram occupy the expected color space in F115W-F444W vs. F200W-F444W one, although photometric scatter may move them outside the color selection window. For this reason, and given also the differences in depths among the various bands, the two diagrams are not expected to select coincident samples in the z$\sim 9-11$ range.

For both color selections described above, we consider only objects with SNR(F444W)$>$8, corresponding to F444W$\sim$28.2, that given the relevant depths in the bluer bands enables the measurement of the large color terms ($\sim$1.7-2) required to separate high-redshift galaxies from the bulk of contaminants. While the proposed diagrams efficiently select high-redshift targets, some contamination from low-redshift galaxies is evident from the redshift distribution of the selected objects (right panels in Fig.~\ref{figcolsel}). The impact of such contamination and the completeness of the selection criteria will be discussed in Sect.~\ref{Sect:complpurity}. Instead, contamination from brown dwarfs is not a concern for our z$\sim$9-15 selection. We show in Fig.~\ref{figcolsel} late type stars with 26.0$\leq$F444W$\leq$28.0 that meet the above mentioned SNR criteria in the F150W to F444W bands: the brown dwarfs occupy a distinct region of the color diagrams.

\begin{figure*}[ht!]
\includegraphics[width=9cm]{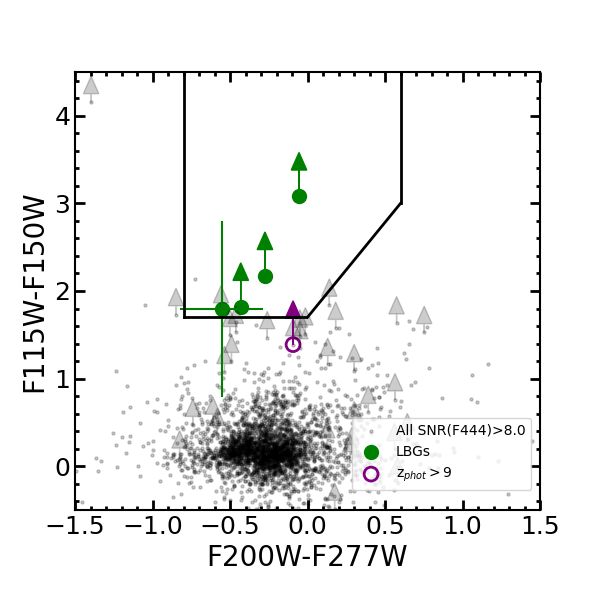}
\includegraphics[width=9cm]{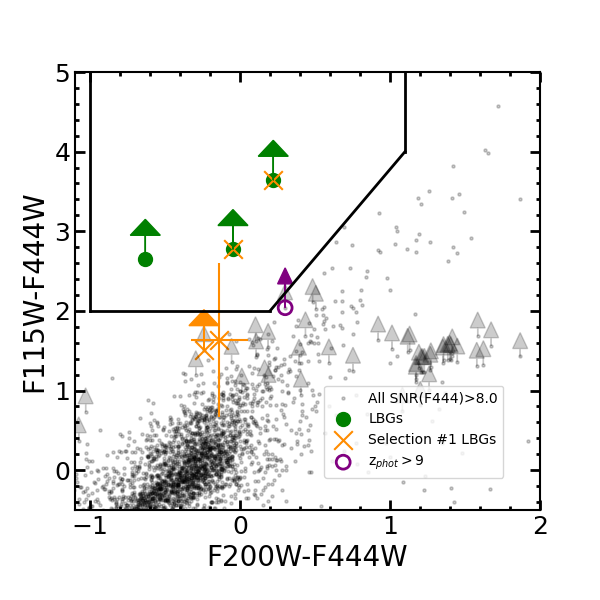}
\caption{Observed color selection diagrams for LBGs at z$\sim$9-11 (left, Selection \#1) and z$\sim$9-15 (right, Selection \#2) in GLASS-JWST. Grey points show all objects with SNR(F444W)$>$8 in the GLASS catalog. Green circles indicate the color-selected candidates. The additional candidate selected on the basis of photometric redshift is shown as a purple empty circle. The z$\sim$9-11 LBGs from the Selection \#1 diagram are shown as dark orange crosses in the Selection \#2 one.  Upper limits are indicated by arrows. All error-bars and upper limits are at 1$\sigma$.\label{figcolselobs}}
\end{figure*}

\subsection{Photometric redshift selection}\label{Sect:photozsel}
In addition, we search for high-redshift candidates that do not meet the color selection criteria defined above but have robust photometric redshift in the range z=9-15. We estimate photometric redshifts with two different codes, \textsc{EAzY} \citep{Brammer2008}, and \textsc{zphot} \citep{Fontana2000}. The \textsc{EAzY} code was run with the default V1.3 spectral template and a flat prior \citep[see the companion paper by][]{Leethochawalit2022}. The analysis with \textsc{zphot} has been performed by fitting the observed photometry with BC03 templates \citep[see the companion paper by][]{Santini2022}. As described by \citet[][]{Merlin2021}, the BC03 library includes templates with both declining and delayed star-formation histories and models the contribution from nebular continuum and line emission following \citet{Schaerer2009} and \citet{Castellano2014}. 
 
Conservatively, photometric redshift candidates are chosen as objects with both $z_{Eazy}>9$ and $z_{zphot}>9$.

We also require SNR(F444W)$>$8 and SNR$>$3 in F200W, F277W and F356W, to ensure a proper measurement of the spectral energy distributions (SEDs). 
Although some true positives may be lost with this approach, the multiple requirements help minimize the risk of contamination.

\subsection{Completeness and Purity}\label{Sect:complpurity}

We used simulations to estimate the completeness of our selection of z$>$9 galaxies. First, simulated NIRCam images in all relevant filters are built out of raw frames obtained by the \textsc{Mirage}\footnote{\url{https://www.stsci.edu/jwst/science-planning/proposal-planning-toolbox/mirage}} software on the basis of the actual scheduling of the Program as derived from the APT. The simulations as described by M22 were used to fine tune the image reduction pipeline, and accurately reproduce the noise properties of the real NIRCam images.

Second, we inserted in blank regions of the simulated images 2.5$\times$10$^5$ mock LBGs at 9$<z<$15 and with a constant distribution at -18.5$<$M$_{UV}<$-22.5. The observed magnitudes are obtained by randomly associating a BC03 model from the library described in Sec.~\ref{Sect:colorsel}. We assume that objects follow a circular \citet{Sersic1968} light profile with n=1 and that their effective radius scales with L$_{UV}$ as r$_{e} \propto L^{0.5}$, consistent with several estimates at comparable redshifts \citep[e.g.][]{Grazian2012,Kawamata2018,Bouwens2022,Yang2022}. Following \citet{Yang2022} we assume an effective radius of 0.8 kpc for objects with M$_{UV}$=-21. In order to avoid overcrowding, simulations are repeated by inserting 500 objects each time. Detection and photometry on the simulated galaxies is performed in the same way as for the real catalog. 

We estimate a completeness of $\sim$70\% ($\sim$60\%) for the selection of bright M$_{UV}<$-20 objects at z$\sim$9-11 (z$\sim$11-15), decreasing to $\sim$60\% ($\sim$50\%) at the nominal limits of M$_{UV}<$-19.0 (-19.5) quoted above. We also find a non negligible chance of detection of fainter targets due to photometric scatter, with a completeness of $\sim$50\% ($\sim$35\%) for the selection of objects of M$_{UV}\sim$-18.5 (-19.0) at z$\sim$9-11 (z$\sim$11-15).

We evaluate potential contamination with two different mock catalogs perturbed with realistic noise reproducing the depth of our data. As a first test, we use the mock catalog described in Sec.~\ref{Sect:colorsel} based on EGG and known UV LFs from the literature. We find that each of our color criteria would select $\sim$0.6 interlopers on an area equivalent to ours.  As expected \citep[e.g.,][]{Vulcani2017}, the interlopers are mostly faint (F444W$\sim$28), red galaxies at z$\sim$2-3 (see Fig.~\ref{figcolsel}).
As a second test we use the mock catalogs from the JAdes extraGalactic Ultradeep Artificial Realizations (JAGUAR) \citep[][]{Williams2018}, including predicted NIRCam fluxes for objects at 0.2$<z<$15 and stellar mass log(M/M$_{\odot}$)$>$6. JAGUAR provides a complementary test with respect to EGG also thanks to the inclusion of emission lines in the predicted SEDs. The potential contamination from objects in the JAGUAR catalog turns out to be much lower than in the previous case, i.e. $<$0.1 objects per selection, per field.
In conclusion, both tests suggest that residual contamination should be small, albeit with some uncertainty. This is inevitable, considering that the 1-5 $\mu$m spectra of intermediate galaxies at these depths are uncharted territory.

\section{Candidate galaxies at $z=9-15$}\label{Sect:sample}
The resulting observed color-color diagrams are shown in Fig.~\ref{figcolselobs}. We select five candidates at z$>$9 by combining the samples from the two selections described in Sec.~\ref{Sect:colorsel}. One additional candidate is selected from the photometric redshift analysis described in Sec.~\ref{Sect:photozsel}. The positions and magnitudes of the candidates are listed in Table~\ref{tab:candidates}. All objects are selected with both the 2FWHM and the $0\farcs2$ aperture colors, with the exception of GHZ5 whose F115W-F150W color is slightly outside the selection boundary when using the 2FWHM ones.The six selected objects have been visually inspected to ensure that they are not affected by defects such as spurious sources, stellar spikes, etc. We found that GHZ2 is very compact and it is close ($\sim$0.5") to a bright, foreground galaxy that contaminates the total flux measured within the Kron ellipse. We thus estimated its total flux using the \textsc{T-PHOT} software \citep[][]{Merlin2015,Merlin2016a} which has been shown to provide accurate photometry for blended sources. We used the GHZ2 light profile in the F277W band, which provides the best compromise between resolution, SNR and contamination from the low redshift neighbour, as high-resolution prior to re-extract the F444W photometry. We found a total magnitude F444W=27.21 $\pm$ 0.04. Finally, we have used \textsc{GALFIT} \citep{Peng2010} to fit its light profile in both the F200W and F444W bands in order to measure its spatial extent and to further assess the measurement of the total flux under different assumptions. We used a \citet[][]{Sersic1968} model and fixed the index $n$ at different values from n=0.5 to n=4. We found that GHZ2 is compact but resolved, with effective radius $1.0\pm0.5$ pixels. The n=1 fit provides the best residuals. We found from \textsc{GALFIT} a total F444W magnitude of F444W=27.25 $\pm$ 0.20, in agreement with the \textsc{T-PHOT} measurement but with a larger error due to the systematic uncertainty on the profile parameters. We adopt this conservative error estimate in the forthcoming analysis.

\begin{figure*}[ht!]
\includegraphics[width=9cm]{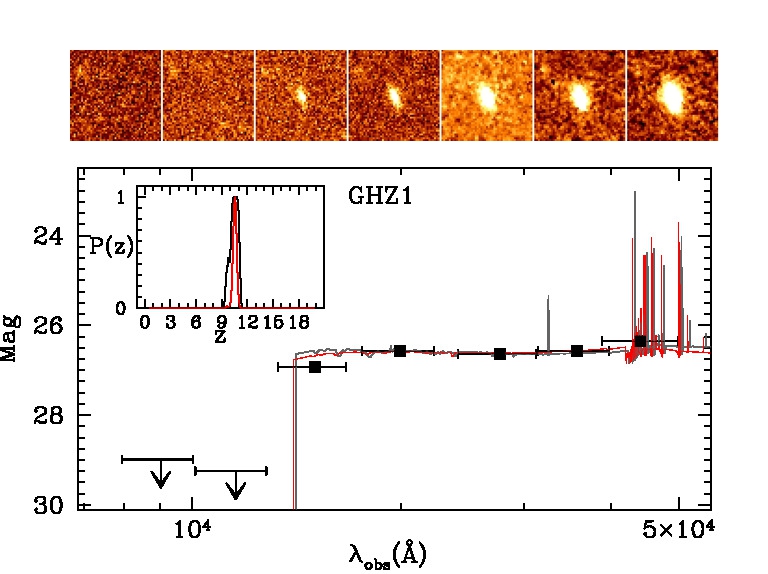}
\includegraphics[width=9cm]{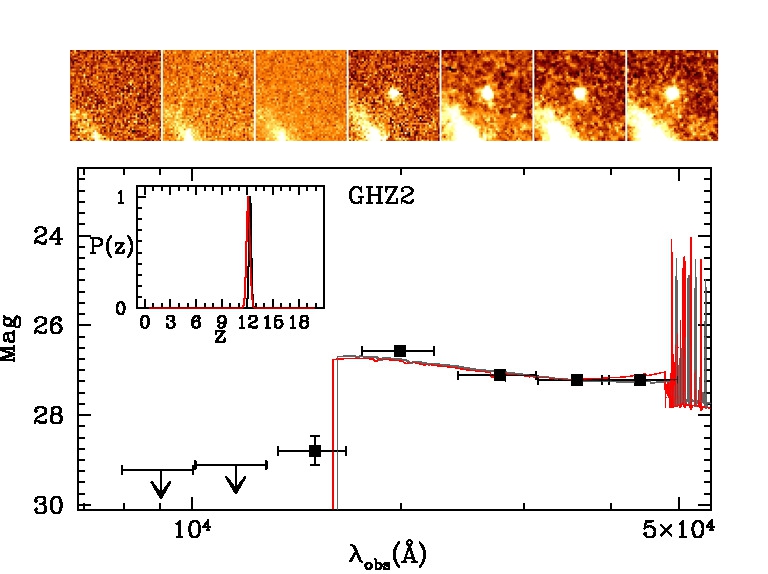}
\caption{The two high-quality bright high-redshift candidates from the GLASS-JWST NIRCAM field taken in parallel to NIRISS. Photometry and best-fit SEDs at the best-fit redshift are given in the main quadrant. Redshift probability distributions P(z) from  \textsc{zphot} (grey) and \textsc{EAzY} (red) are shown in the inset. Upper limits are reported at the $2\sigma$ level, including a conservative estimate of the error budget, especially in the bluest bands (M22). Thumbnails, from left to right, show the objects in the F090W, F115W, F150W, F200W, F277W, F356W and F444W bands. \label{highz1}}
\end{figure*}

\begin{figure*}[ht!]
\includegraphics[width=9cm]{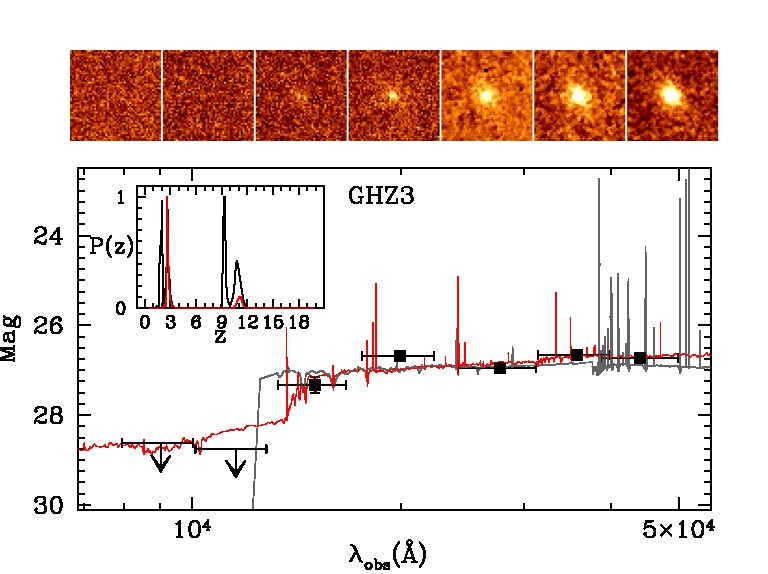}
\includegraphics[width=9cm]{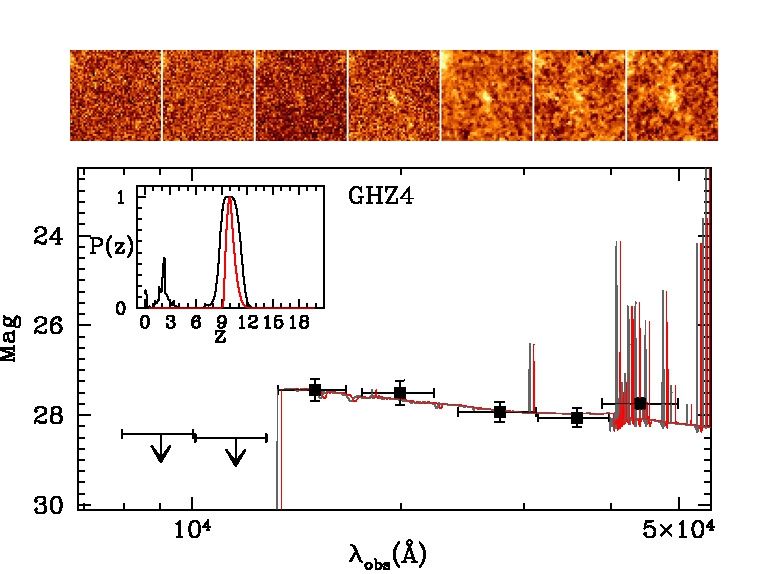}
\includegraphics[width=9cm]{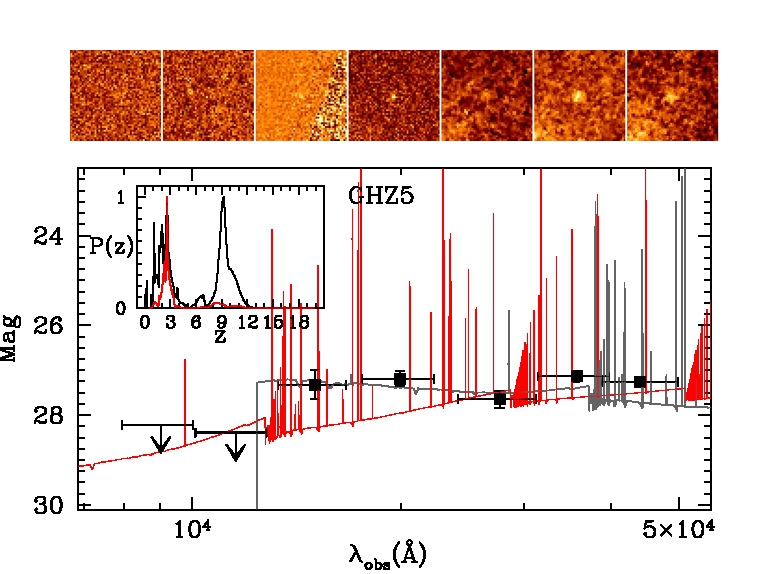}
\includegraphics[width=9cm]{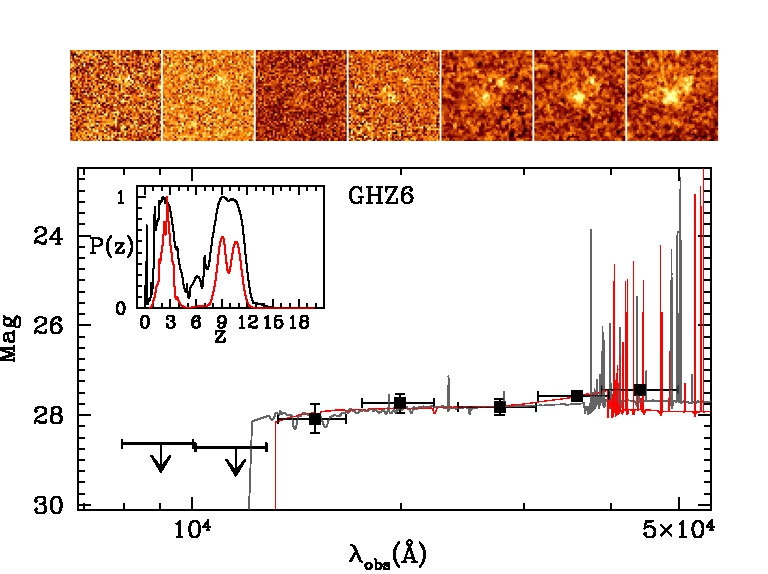}
\caption{As in Figure~\ref{highz1}, for fainter high-redshift candidates from GLASS-JWST. These candidates need further verification owing to the non-negligible probability of a low-redshift solution. \label{highz2}
}
\end{figure*}

Table~\ref{tab:candidates} also reports the UV rest-frame magnitudes estimated from the observed F200W at the median photometric redshift, the UV slope obtained by a fit on the observed F200W, F277W and F356W bands, the effective radius \citep[][]{Yang2022b}, and the SFR from the SED-fitting performed with \textsc{zphot} \citep[][]{Santini2022}.
We show in Fig.~\ref{highz1} and Fig.~\ref{highz2} the best-fit SEDs, the PDF(z) and thumbnails for all candidates.

We highlight the detection of the two bright sources GHZ1 and GHZ2, at $z\simeq 10.6$ and $z\simeq12.2$, respectively. These objects are located in the color-color diagram fairly away from color selection borders, and have no significant low-redshift solution. Their P(z) is single peaked, the 68\% confidence level photometric redshift range being $z_{zphot}=$9.68-11.15 ($z_{EAzY}=$10.28-10.80) for GHZ1 and  $z_{zphot}=$12.03-12.58 ($z_{EAzY}=$11.85-12.39) for GHZ2.  They are among the most robust candidates at $z>9$ ever detected, as demonstrated by the several recent works that have presented their selection with redshift and properties comparable to those discussed here \citep[][]{Donnan2022,Harikane2022b,Naidu2022b}. They are remarkably bright (M$_{UV}\lesssim$-21) and star-forming (SFR$\gtrsim$20$M_{\odot}$/yr). GHZ2 is  compact with R$_e$=120$\pm$10 pc \citep[][]{Yang2022b} and it has a very steep UV slope $\beta$=-3.0. This extreme value is partly driven by the F200W band, as the slope value is $\beta$=-2.4$\pm$0.2 when using the F277W and F356W bands only. These UV slope estimates are anyway indicative of very young ages and very low metallicity, possibly pristine, stellar populations which will need to be confirmed through JWST spectroscopy. A third bright candidate (GHZ3, M$_{UV}\sim$-20.7, SFR$\sim$30$M_{\odot}$/yr) is also selected by both of our color diagrams, but its photometric redshift is more uncertain: it is found to be at z$\sim$9.3 by \textsc{zphot}, while \textsc{EAzY} gives a best-fit solution at z$\sim$2.7. The remaining candidates are fainter (M$_{UV}\gtrsim$-20.2) galaxies likely at z$\sim$9-10.

The GHZ3-6 candidates exhibit in general alternative lower-z solutions with non-negligible probability. These lower-z solutions are driven in part by our conservative error estimates, especially in the bluer bands, accounting for data reduction and calibration systematics (M22). However, definitive confirmation will require deep imaging at shorter wavelengths. Instead, contamination from galactic stars can be excluded because all objects show resolved profiles as discussed by \citet[][]{Yang2022b} \footnote{Reassuringly, the candidate T-dwarf star GLASS-JWST-BD1 found in the GLASS-JWST-ERS field by \citet[][]{Nonino2022} is not selected by any of our criteria.} 
 
The physical properties, morphology, and size of our candidates are discussed in detail in the companion papers by \citet[][]{Santini2022}, \citet[][]{Treu2022}, and \citet[][]{Yang2022b} respectively. 

\begin{deluxetable*}{ccccccccccc}\label{tab:candidates}
\tablecaption{Galaxy candidates at z$>9$ in GLASS-JWST--ERS$^a$}
\tablewidth{0pt}
\tablehead{
\colhead{ID} & \colhead{R.A.} & \colhead{Dec} & \colhead{F444W} & \colhead{z$_{EAzY}$} & \colhead{z$_{zphot}$}  & \colhead{M$_{UV}$} &  \colhead{$\beta$}  & \colhead{SFR} &  \colhead{R$_{e}$}  &\colhead{Selection}\\
\colhead{} & \colhead{deg.} & \colhead{deg.} & \colhead{} & \colhead{} & \colhead{}  & \colhead{} & \colhead{} & \colhead{$M_{\odot}$/yr}&  \colhead{kpc}  & \colhead{}
}
\startdata
GHZ1 &  3.511929 & -30.371848 &  26.36$\pm$0.05 &  10.53 &  10.63& -20.98$\pm$0.06 & -1.99$\pm$0.10 &25$^{+68}_{-16}$ &   0.43 $\pm$ 0.02 & 1,2 \\
 GHZ2$^b$ &  3.498985 & -30.324771 &  27.21$\pm$0.20 &  12.11 &  12.30& -21.19$\pm$0.20 & -3.00$\pm$0.12 & $20^{+14}_{-13}$&0.12 $\pm$ 0.01  &2\\
 GHZ3 &  3.528937 & -30.363811 &  26.73$\pm$0.07 &   2.69 &   9.33& -20.69$\pm$0.09 &-1.85$\pm$0.17 & 31$^{+10}_{-8}$& 0.88 $\pm$ 0.09  &  1,2 \\
 GHZ4 &  3.513743 & -30.351554 &  27.74$\pm$0.12 & 10.08 &   9.93& -19.98$\pm$0.27 & -2.86$\pm$0.55 & 5$^{+19}_{-3}$& 0.39 $\pm$ 0.09  &  1,2 \\
 GHZ5 &  3.494437 & -30.307620 &  27.25$\pm$0.08 &   2.49 &   9.20& -20.17$\pm$0.18 & -1.82$\pm$0.33& 18$^{+21}_{-15}$& 0.21 $\pm$ 0.08 & 1 \\
 GHZ6 &  3.479054 & -30.314925 &  27.43$\pm$0.11 &   9.85 &   9.05&-19.66$\pm$0.21 & -1.67$\pm$0.38 &  10$^{+30}_{-8}$&0.45 $\pm$ 0.09 &photo-z\\
\enddata
\tablecomments{ a) Coordinates and fluxes are from the GLASS-JWST catalog by M22, the rest-frame M$_{UV}$ has been obtained by converting the F200W magnitude at the average photometric redshift, except for GHZ3 and GHZ5 where the only solution at z$>9$ was used. The UV slope $\beta$ is obtained by fitting the F200W, F277W and F356W bands, the uncertainties in the fit account for photometric errors \citep[][]{Castellano2012}. The R$_{e}$ has been estimated from the F200W image by \citet[][]{Yang2022b}. Physical parameters from \textsc{zphot} \citep[][]{Santini2022}. The last column indicate selections \#1 and \#2 in Sect.~\ref{Sect:colorsel}, and additional photo-z candidates not meeting color selection criteria. b) All errors are computed considering the systematic uncertainty from different photometric techniques (Sect.~\ref{Sect:sample}).}
\end{deluxetable*}

\section{Summary and Discussion}

We have presented a search for galaxy candidates at z$\sim$9-15 from the seven-bands, mag $\sim$29 (5$\sigma$) NIRCam imaging observations, acquired by the GLASS-JWST Early Release Science Program in parallel with NIRISS primary observations of the cluster A2744. The NIRCam images enable the first multi-band selection of Lyman-break candidates at z$\sim$9-15 through two independent color-color diagrams.

We have identified two bright ($M_{UV}\simeq -21$) sources which, thanks to the depth of our images, are unambiguously placed  at $z=10.6$ and $z\simeq 12.2$, respectively. These are among the most reliable candidates ever selected at $z>9$ via photometric techniques.

We have also identified five other fainter candidates at $z>9$. Some of these objects have significant alternative solutions at lower redshift, that it is difficult to exclude given the limits that are currently achieved by our data, and residual systematic uncertainties especially in the bluest bands. Definitive confirmation will require deeper images at short wavelengths as well as NIRSpec spectroscopic follow-up.

Keeping in mind the uncertainties associated with our limited area and to the presence of a foreground cluster that may boost the luminosity of our candidates, it is intriguing to explore the potential implications of our results on the evolution of galaxies at large redshift.

Noting that the F444W$\sim$28.2 limit corresponds to M$_{UV}\lesssim$-19. (-19.5) at $z=10 (13)$, we estimate the average number of sources expected in our field under several assumptions about the evolution of the LF.

At z$\sim$9-11 we expect 1.7$_{-0.6}^{+1.1}$ on the basis of the z$\sim$9 and z$\sim$10 UV LFs by \citet[][]{Bouwens2021} and \citet[][]{Oesch2018}, respectively,  where the uncertainties consider the relevant 68\%c.l. confidence range in the LF parameters. The double power law (DPL) model by \citet[][]{Bowler2020} predicts 2.4$_{-1.9}^{+1.6}$ galaxies. We expect as many as 6.8$_{-1.8}^{+2.9}$ objects at z$\sim$9-11 on the basis of the model by \citet[][]{Mason2015}. 

At z$>$11, where no direct constraints exists, we can estimate the average number of sources with M$_{UV}<$-19.5 under four scenarios. In the first (optimistic) case we assume that the $z=10$ UV LF remains constant up  to $z=15$ and predict 0.6$_{-0.3}^{+0.5}$ objects in one GLASS-JWST-ERS parallel field. A perhaps more realistic scenario is provided by the evolving UV LF by \citet[][]{Bouwens2021} that extrapolates at z$>$11 the observed evolution in the \citet[][]{Schechter1976} parameters of the UV LFs measured at lower redshifts. In this second scenario, on average, only 0.1$_{-0.05}^{+0.1}$ objects fall within one NIRCAM parallel field. A similar number of 0.08$_{-0.03}^{+0.06}$ objects is predicted based on \citet[][]{Mason2015}. Finally, the extrapolation at z=11-15 of the evolving DPL model by \citet[][]{Bowler2020} predicts 0.5$_{-0.2}^{+0.3}$ objects.

We note that such predictions  are affected by cosmic variance at the level of $\sim$40\%\footnote{Following the Cosmic Variance Calculator at https://www.ph.unimelb.edu.au/~mtrenti/cvc/CosmicVariance.html. See \citet[][]{Trenti2008}.}.
We also note that the adopted M$_{UV}$ limits may be altered by photometric scatter. In combination with the steep increase in the number counts beyond the knee of the LF, a non negligible number of fainter sources can thus be scattered  within the observable range. As a reference, by adopting 0.5 mag fainter M$_{UV}$ cuts, the predicted number of objects at z$\sim$9-11 increases to 3.5 on the basis of the observed UV LFs, and to 7.5 in the \citet[][]{Mason2015} model. The predicted number of sources at z$>$11 increases to 1.5 (0.2)  assuming a constant (evolving) UV LF as above, and to 0.26 on the basis of \citet[][]{Mason2015}. 

Considering the incompleteness of our sample, as estimated above, we conclude that the total number of detected sources is roughly in line  with the expectations of a non-evolving LF. 

What is most remarkable of this first search is that we have found two bright (M$_{UV}\lesssim$-21) sources, one at $z\sim10$ and one at z$\sim$12, well beyond the expectations based on the extrapolation of the LF at lower redshift. In fact, all observations and models predict a negligible number of sources brighter than this limit in the redshift range z=9-15 on an area equal to the GLASS one: the predictions being of $\lesssim$0.2 objects, compared to an observed number of 2$^{+2.7}_{-1.4}$ \citep[where the uncertainty includes both cosmic variance and the Poisson uncertainty for small number of events computed following][]{Gehrels1986}. The aforementioned predictions are summarised in Table~\ref{tab:LFs}. Clearly it is premature to draw broad conclusion based on a single field, considering small number statistics,  cosmic variance, clustering and lensing. In particular, the impact of lensing on these results will need to be assessed. A preliminary investigation \citep[based on the model by][]{Bergamini2022} suggests that  magnification may be 2-10 times higher than the 10-15\% value measured in the HFF Abell2744 Parallel \citep[][]{Castellano2016b}. At the present stage this indication is highly uncertain because it is extrapolated from the cluster core to the NIRCam field 3 arcmins away and, most importantly, it is not possible to obtain lensing constraints for single sources. We note that our candidates do not show any evidence of being sheared, so although magnification of up to 50\% is possible, it is unlikely to be much higher. A thorough analysis will be performed upon availability of additional spectroscopic information in the GLASS-JWST-ERS field.
Despite the aforementioned sources of uncertainty, these results show that JWST can obtain the scientific results it has been designed for.

\begin{table*}[ht]
\caption{Predictions on the number of z$>9$ objects in GLASS-JWST--ERS$^a$}
\centering
\begin{tabular}{|c|c|c|c|c|}
\hline
 \multirow{2}*{UV LF} & \multicolumn{2}{c|}{z=9-11}&\multicolumn{2}{c|}{z$>$11}\\
\cline{2-5}
 & M$_{UV}<$-21.0 & M$_{UV}<$-19.0 &  M$_{UV}<$-21.0 & M$_{UV}<$-19.5 \\ \hline
Oesch+18$^b$ & $<$0.06 & 1.7$_{-0.6}^{+1.1}$ & $<$0.04  & 0.6$_{-0.3}^{+0.5}$\\
\hline
LF(z) Bouwens+21$^c$  & '' &  '' & $<$0.01 &  0.1$_{-0.05}^{+0.1}$\\
\hline
Mason+15$^d$ & $0.16_{-0.05}^{+0.07}$	 & $6.8_{-1.8}^{+2.9}$  & $0.002\pm0.001$ & $0.08_{-0.03}^{+0.06}$\\
\hline
Bowler+20$^e$ & $0.06_{-0.03}^{+0.30}$	 & $2.4_{-1.9}^{+1.6}$  & $0.05_{-0.04}^{+0.1}$ & $0.5_{-0.2}^{+0.3}$\\
\hline
\end{tabular}\label{tab:LFs}
\small \\$^a$ Uncertainties on the predicted numbers are based on the 68\%c.l. errors on the Schechter parameters; $^b$ UV LF at z$\sim$10 by \citet[][]{Oesch2018}, assumed fixed at z$>$11; $^c$ Extrapolation at z$>$11 of the UV LF as a function of redshift by \citet[][]{Bouwens2021}. $^d$ The semi-empirical UV LF model by \citet{Mason2015}. $^e$ Extrapolation of the DPL model as a function of redshift by \citet[][]{Bowler2020}.
\end{table*} 

There is no doubt that the  deeper and/or wider surveys already planned will eventually gather the long-awaited-for sample of galaxies in the reionization epoch and revise our understanding of how and when bright galaxies formed. 

This first JWST-based search for candidate galaxies also provides us with compelling targets for spectroscopic follow-up in JWST Cycle-2. In addition to providing the necessary redshift confirmation, the relatively high luminosity of these galaxies means that NIRSpec spectroscopy should enable the study of their physical properties such as star formation rate, dust content, and gas metallicity.

\begin{acknowledgments}
This work is based on observations made with the NASA/ESA/CSA James Webb Space Telescope. The data were obtained from the Mikulski Archive for Space Telescopes at the Space Telescope Science Institute, which is operated by the Association of Universities for Research in Astronomy, Inc., under NASA contract NAS 5-03127 for JWST. These observations are associated with program JWST-ERS-1324. The JWST data used in this paper can be found on MAST: http://dx.doi.org/10.17909/fqaq-p393. We acknowledge financial support from NASA through grants JWST-ERS-1342. KG and TN acknowledges support from Australian Research Council Laureate Fellowship FL180100060. CM acknowledges support by the VILLUM FONDEN under grant 37459. The Cosmic Dawn Center (DAWN) is funded by the Danish National Research Foundation under grant DNRF140. We acknowledge support from INAF "Minigrant" "Reionization and fundamental cosmology with high-redshift galaxies".
We thank Pascal Oesch and Rohan Naidu for the useful feedback and discussion.
\end{acknowledgments}

%

\vspace{5mm}


\software{Astropy \citep{Astropy2013}, EAzY \citep{Brammer2008}, EGG \citep{Schreiber2017}, Matplotlib \citep{Hunter2007}, SExtractor \citep[v2.8.6][]{Bertin1996,Guo2013}, T-PHOT \citep[][]{Merlin2015,Merlin2016a}, zphot \citep{Fontana2000}}





\bibliography{biblio2}{}
\bibliographystyle{aasjournal}



\end{document}